# Multimodal Spatial Attention Module for Targeting Multimodal PET-CT Lung Tumor Segmentation

Xiaohang Fu, Lei Bi, Ashnil Kumar, Michael Fulham, and Jinman Kim

*Abstract*—**Multimodal positron emission tomography-computed tomography (PET-CT) is used routinely in the assessment of cancer. PET-CT combines the high sensitivity for tumor detection with PET and anatomical information from CT. Tumor segmentation is a critical element of PET-CT but at present, there is not an accurate automated segmentation method. Segmentation tends to be done manually by different imaging experts and it is labor-intensive and prone to errors and inconsistency. Previous automated segmentation methods largely focused on fusing information that is extracted separately from the PET and CT modalities, with the underlying assumption that each modality contains complementary information. However, these methods do not fully exploit the high PET tumor sensitivity that can guide the segmentation. We introduce a multimodal spatial attention module (MSAM) that automatically learns to emphasize regions (spatial areas) related to tumors and suppress normal regions with physiologic high-uptake. The resulting spatial attention maps are subsequently employed to target a convolutional neural network (CNN) for segmentation of areas with higher tumor likelihood. Our MSAM can be applied to common backbone architectures and trained end-to-end. Our experimental results on two clinical PET-CT datasets of non-small cell lung cancer (NSCLC) and soft tissue sarcoma (STS) validate the effectiveness of the MSAM in these different cancer types. We show that our MSAM, with a conventional U-Net backbone, surpasses the state-of-the-art lung tumor segmentation approach by a margin of 7.6% in Dice similarity coefficient (DSC).**

*Index Terms*—**Convolutional Neural Network (CNN), Multimodal Image Segmentation, PET-CT**

## I. INTRODUCTION

ACCURATE tumor delineation in patients with cancer is necessary for effective diagnosis, treatment planning, radiomics analysis, and personalized medicine [1]. The integrated imaging modality PET-CT, which combines positron emission tomography (PET) and computed tomography (CT), is increasingly the modality of choice for a number of cancers including non-small cell lung cancer (NSCLC) [2]. PET-CT leverages the functional nature of PET with its high sensitivity for detecting abnormal tumor metabolism to improve the diagnosis, staging, and assessment of tumors over the anatomical limitations of CT alone, where alterations in size are required to identify disease [1]. $^{18}$F-fluorodeoxyglucose ($^{18}$F-FDG) is the most common PET radiopharmaceutical used in oncological PET, and tumors are usually readily identified as regions of high FDG uptake or 'hot-spots' [3]. The degree of FDG uptake can be semi-quantified by using the standard uptake value (SUV), defined as the ratio of radioactivity concentration in the region of interest (ROI) to the concentration in the body [4].

Despite the obvious value of having a functional parameter of tumor activity that is detected with PET, PET-CT tumor segmentation is challenging. In PET, it is difficult to determine the spatial extent of the tumor as PET images have poor resolution when compared to CT [5]. Further, normal structures - the heart, bladder, and brown fat - and benign processes including inflammation, can display varying degrees of increased FDG uptake [6]. Thus, at times it can be difficult to determine if focal regions of increased FDG uptake are related to tumors from PET alone. Hence, PET images are always interpreted with the corresponding CT image [7]. As such, PET-CT tumor segmentation still relies upon specialist imaging expertise to discern the relevant information captured by each modality with attendant high costs, and inter- and intra-observer inconsistencies [8, 9]. Automated PET-CT tumor segmentation is uniquely challenging due to the additional complexity of needing to consider the complementary features from each modality. The optimal extraction and application of the data from PET and CT is a relatively under-studied topic compared to unimodal imaging problems and robust methodologies are much anticipated [10].

Various strategies for automatic PET-CT segmentation have been proposed. These include thresholding [11-13], which aims to separate tumors from the background based on SUV differences. A wide range of SUV thresholds have been used in the clinical setting including an SUV of >2.5, or 41% to 90% of the maximum value to identify a tumor [5]. Thresholding, however, can be flawed because some normal physiological processes and benign conditions such as pneumonia can have very high FDG uptake and some primary tumors can have SUV<2.5. In addition, the type of scanner used, the time

X. Fu, L. Bi, A. Kumar, and J. Kim are with the School of Computer Science, Faculty of Engineering, The University of Sydney, NSW 2006, Australia (e-mail: xiaohang.fu@sydney.edu.au, lei.bi@sydney.edu.au, ashnil.kumar@sydney.edu.au, and jinman.kim@sydney.edu.au).

M. Fulham is with the Department of Molecular Imaging, Royal Prince Alfred Hospital, Australia, and with the Sydney Medical School, The University of Sydney, Australia (e-mail: michael.fulham@sydney.edu.au).

Corresponding author: Jinman Kim.

between the injection of the FDG and the commencement of data acquisition (the uptake period), image reconstruction method, the calculation of SUV by the scanner vendor, image noise, and tissue of interest can all affect the SUV. Hence threshold selection requires specialist domain knowledge of PET-CT imaging [5]. Thresholding-based methods have generally been superseded as the limitations have been identified, and as computational power and techniques continue to advance [12].

Other strategies that have been explored include the fusion of modality-specific features or complementary information from PET and CT, including graph-based methods [7, 14-18]. Han et al. [16] formulated the tumor segmentation problem as a graph-based Markov Random Field (MRF) with an energy function that used advantageous characteristics of each modality and penalized the segmentation difference between PET and CT images. Bagci et al. [7] proposed a random walk method for co-segmentation of multiple objects in PET, PET-CT, MRI-PET, and fused MRI-PET-CT images via a hyper-graph. Other methods such as stochastic modeling [19], active contours [20], co-clustering and belief functions [21] have also been used. Other investigators used one modality to guide tumor localization in another modality. Wojak et al. [22] proposed a joint variational segmentation method using PET intensities to provide local constraints to adjust the segmentations on CT. Bagci et al. [15] proposed a random walk co-segmentation method that thresholds FDG uptake values in PET to automatically initialize foreground and background seeds, and then finds corresponding boundaries in the CT image. These methods which used PET to drive segmentation, however, did not consider the spatial and contextual characteristics of the PET image, as contours were only computed on the CT. Further, they are highly dependent on the PET SUVs, so are inherently limited in the presence of normal high-uptake activity.

State-of-the-art automated segmentation methods are now typically based on deep learning (DL). With medical images, various convolutional neural networks (CNNs), especially U-Net [23], have proven valuable across a wide range of segmentation problems. This success can partly be attributed to the ability of DL methods to automatically learn to extract features from images that are meaningful to the task at hand. Recently, a number of investigators have reported on DL approaches for PET-CT tumor segmentation. Li et al. [24] processed CT with a fully convolutional network (FCN), and PET with a fuzzy variational model, then integrated the probability maps from the models. Zhong et al. [25] used graph-based co-segmentation to combine outputs from two separate 3D U-Nets for each modality. Strategies that fuse features at various points within CNNs have also been used [26-28]. Rather than combining feature volumes with a simple addition or concatenation operation without consideration of spatial context, Kumar et al. [28] proposed a CNN model that automatically learns the relative spatial importance of each modality's features to prioritize content from PET or CT at different locations, and then fuses the weighted features. In general, however, these recent approaches do not fully exploit the sensitivity of PET. During conventional manual analysis, hotspots on PET images draw the attention of experts to the corresponding locations in the CT scan, which are analyzed to determine if the pixels in the PET image correspond to disease or a benign process [6, 29]. Hence for this work, we developed a spatial attention module that exploits the high sensitivity of PET to enhance tumor segmentation in PET-CT data.

Attention mechanisms that extract and highlight salient information and minimize irrelevant features with regards to the problem context have proven valuable in DL applications in computer vision [30-35]. To date, an attention approach has not been designed for PET-CT, which is relatively unique in that one modality (PET) is more important in directing attention toward the tumor. Our proposed multimodal spatial attention module (MSAM) can be integrated into and trained end-to-end via standard backpropagation with a backbone CNN architecture without additional supervision or domain knowledge. The MSAM automatically learns to differentiate high-uptake normal and abnormal tumor regions on PET, increases the focus on tumor regions, and decreases the influence of irrelevant regions to enhance PET-CT tumor segmentation performance.

Our contribution to current segmentation approaches are as follows: a) we introduce a DL attention subnetwork module in multimodal PET-CT image analysis; b) we use attention maps derived from PET data to focus a segmentation CNN to areas of the CT image that have greater tumor likelihood; c) we compare our approach against current spatial attention methods and demonstrate that our approach exposes tumor regions with superior visual clarity, and provides greater improvement to the segmentation performance of a backbone CNN.

## II. METHODS

### A. Materials

We used two PET-CT datasets – one from patients with NSCLC and one from patients with soft tissue sarcomas (STSs). The NSCLC dataset comprised of 50 patients with pathologically proven NSCLC, acquired on a Biograph 128-slice mCT (PET-CT scanner; Siemens Healthineers, Hoffman Estates, Il, USA). The original CT and PET image resolutions were $512 \times 512$ at $0.98$ mm $\times 0.98$ mm for CT and $200 \times 200$ at $4.07$ mm $\times 4.07$ mm for PET. The interslice distance (slice thickness) for CT and PET volumes was 3 mm. Tumor regions were delineated using a semi-automatic process which involved localizing the primary tumor and any involved thoracic lymph nodes by an experienced imaging specialist. Connected thresholding was then applied to extract the tumor regions, followed by minor manual adjustments where necessary to improve the segmentation. The resulting annotations were used as the ground truth for evaluation. The STS dataset was a public dataset comprising FDG PET-CT and magnetic resonance imaging (MRI) scans from 51 patients with histologically proven STSs [36]. The FDG PET-CT scans were acquired on a Discovery ST scanner (GE Healthcare, Waukesha, WI). The slice thickness of all PET volumes was 3.27 mm, with a median

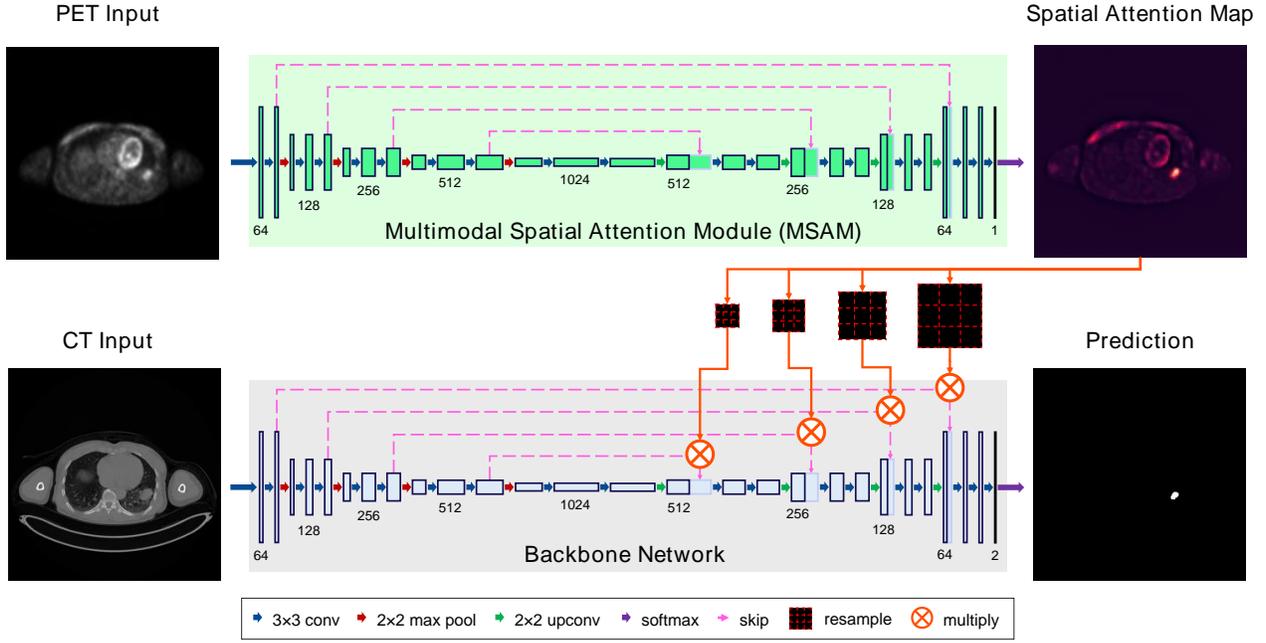

**PET Input**

**Spatial Attention Map**

Multimodal Spatial Attention Module (MSAM)

**CT Input**

**Prediction**

Backbone Network

3×3 conv • 2×2 max pool • 2×2 upconv • softmax → skip ■ resample ⊗ multiply in-plane resolution of $5.47$ mm $\times\ 5.47$ mm (range: $3.91$–$5.47$ mm). The tumor contours were manually delineated by an expert radiation oncologist.

Fig. 1. Schematic of our MSAM (green shading): MSAM is integrated with a general CNN-based segmentation model (gray shading); output is a single channel spatial attention map, which is resized to and multiplied elementwise with skip connections between the encoder and decoder of the CNN.

All images from both datasets used in our experiments were rescaled to $256 \times 256$ in lateral resolution. PET image intensities were converted to SUVs, and CT image intensities were converted to Hounsfield units. For our experiments, we exclude slices without tumor pixels in the ground truth.

### B. Overview of our Proposed Method

Our proposed model consists of two main components: the MSAM subnetwork and an encoder-decoder backbone CNN. The MSAM processes the input PET image to infer a spatial attention map to guide tumor localization. The backbone extracts tumors from the CT data. The spatial attention map from PET is then applied to the CT feature maps produced by different scales (stages) of the segmentation backbone. The CT feature maps are thereby focused onto the areas with the strongest spatial attention from PET to produce the final segmentation. As such, the network uses spatial information from both CT and PET in a way that takes advantage of the strengths of each modality. Our model is illustrated in Fig. 1.

### C. Multimodal Spatial Attention Module (MSAM) Design

The MSAM was designed to be a subnetwork that learns to produce an attention map $\boldsymbol{M} \in \mathbb{R}^{H \times W \times 1}$ from an input PET image $\boldsymbol{P} \in \mathbb{R}^{H \times W \times 1}$:

$$\boldsymbol{M} = \text{MSAM}(\boldsymbol{W}, \boldsymbol{b}; \boldsymbol{P}) \qquad (1)$$

where $\boldsymbol{W}$ are the weights of the convolutional layers of the

MSAM and $\boldsymbol{b}$ are the biases. The use of Rectified Linear Unit (ReLU) [37] activation functions within the MSAM means that in any given feature map, elements that have negative feature values are eliminated. That is, the value at coordinates $(x, y)$ will be given by:

$$M(x,y) = \begin{cases} wP(x,y) + b \\ 0 \end{cases} \qquad (2)$$

where $w \in \boldsymbol{W}$ and $b \in \boldsymbol{b}$. This formulation of the map, when incorporated with the segmentation backbone, emphasizes areas of high saliency with large values, and linearly scales down the contribution of areas with less importance.

We used the well-established U-Net [23] for the MSAM subnetwork. The learnable convolutional weights and max pooling components of the U-Net encoder highlight salient areas of the input PET image, while de-emphasizing irrelevant components. The upsampling components return the image to its original size to produce the final map.

### D. Backbone Integration and Segmentation

The MSAM can be integrated into any general CNN architecture containing skip connections. To apply $\boldsymbol{M}$ as attention, the map is multiplied with the feature map produced by the corresponding skip layer $\boldsymbol{L} \in \mathbb{R}^{H \times W \times C}$ per channel to produce the gated skip connection $\boldsymbol{G} \in \mathbb{R}^{H \times W \times C}$:

$$\boldsymbol{G} = \boldsymbol{L} \otimes \Psi(\boldsymbol{M}) \qquad (3)$$

where $\otimes$ denotes element-wise multiplication with $\boldsymbol{M}$ being broadcasted channel-wise, and $\Psi$ the resampling function to downsample $\boldsymbol{M}$ where necessary via bilinear interpolation to the lateral resolution of the skip feature layers in the backbone network. Thus, the skip layer activations are weighted spatially to emphasize relevant regions and suppress non-salient areas. *G* is subsequently concatenated with the corresponding feature volume produced by transposed convolution (deconvolution or upsampling) layers in the decoder stream.

The PET-CT scans are processed slice-wise. The output of the entire network is a 2-channel volume of the same lateral resolution as the input images, over which the softmax function is applied channel-wise to produce per-pixel probabilities of tumor and background.

The MSAM parameters train automatically via standard backpropagation alongside the backbone CNN. There is no need for any additional auxiliary loss functions to guide parameter optimization of this module. The loss function that is applied to train the segmentation CNN also applies to the MSAM. In this way, the module learns without needing to infuse specialist domain knowledge or tune extra hyperparameters.

### E. Implementation Details

We kept the following hyperparameter and implementation choices consistent for all our experiments. The networks were trained end-to-end for 100 epochs with a batch size of 4. We employed the Adam optimizer [38] to minimize the mean per-pixel cross-entropy loss at a fixed learning rate of 0.0001, with a first moment estimate of 0.9 and a second moment estimate of 0.999. Convolutional filter weights were initialized using He et al.'s method [39] while biases were initialized to zero. Dropout was not used in any experiment.

Each input image was mean-subtracted and normalized to unit variance using the training set mean and standard deviation of its image type (PET or CT). We employed standard online (on the fly) image data augmentation by randomly applying a flip (horizontal or vertical), or rotation (of 90, 180 or 270 degrees) to each input training triplet (PET, CT, and segmentation). The order of training examples was re-randomized for every epoch. All networks were implemented based on the TensorFlow framework [40]. Both training and testing were performed with a 12GB NVIDIA GTX Titan X GPU. Training took 2 hours for the lung cancer dataset and 7 hours for the STS dataset.

### F. Evaluation Setup

The main baseline segmentation architecture we used was U-Net [23]. We investigated different input combinations into the backbone CNN with and without MSAM to determine the contributions of each modality and MSAM to the segmentation performance. We used three different inputs without MSAM to ascertain baseline performance without attentional mechanisms: CT only, PET only, or channel-wise concatenated PET-CT. We evaluated segmentation using PET in both the backbone and the MSAM, and the proposed combination of CT in the backbone and PET in the MSAM, to understand the contribution of the MSAM. Further, we verified that the proposed input configuration of feeding CT into the backbone and PET into MSAM surpassed the performance of the following alternatives: a) concatenated PET-CT into the backbone and PET into MSAM; b) CT into the backbone and concatenated PET-CT into MSAM and, c) concatenated PET-CT into both the backbone and MSAM.

We also investigated another baseline in which the encoder of U-Net was substituted with ResNet-50 [41], to demonstrate the generalizability of MSAM for different CNN backbones. The final average pooling layer of ResNet-50 is not needed for segmentation and was thus discarded. The MSAM was applied at each skip connection layer in the backbone. We note, however, that this model configuration lacks a skip connection at the resolution level of the full input image, due to the initial $7 \times 7$ convolution layer in ResNet-50.

We compared our MSAM to three recent image spatial attention approaches: attention residual learning (ARL) [31], convolutional block attention module (CBAM) [32], and attention gates (AG) [30]. ARL modifies the canonical residual block in ResNets by generating extra attention weightings from the identity map and the output of the last convolutional layer of each block. ARL is only compatible with networks with residual blocks. The initial value of the learnable weighting factor for the attention maps in ARL blocks was set to 0.001, as used in the original paper [31]. CBAM has two submodules which infer channel-wise and spatial attention for each feature volume, while the AG module produces a spatial attention map from the downsampling and upsampling paths in a CNN to gate the skip connections. The CBAM and AG mechanisms were inserted to gate each of the skip connections in U-Net, as applied likewise for MSAM. For all benchmark attention methods, we used a PET-only input, or a channel-wise concatenation of PET-CT to provide information from both modalities. Following official implementations, batch normalization with a momentum of 0.99 was used in experiments involving ResNet-50 and CBAM. We also compared our segmentation architecture against state-of-the-art PET-CT lung tumor segmentation methods where deep learning is used to combine complementary information from the two imaging modalities [24, 25, 28].

5-fold cross-validation was carried out for each dataset and all methods. The scans were randomly divided into training and testing sets with an 80/20 percent split – 40 patients for training and 10 for testing (for the STS dataset, one scan was randomly excluded so that the number of scans across each fold was the same). Identical patient splits were used for each method and we ensured that no patient was in both the training and test sets of a fold.

Our main performance metric was the Dice similarity coefficient (DSC), which combines precision and sensitivity via a harmonic mean. Since we are interested in tumor segmentation, we only considered the DSC of the tumor regions. We also report precision, sensitivity (equivalent to recall), and specificity scores. All our scores are pixel-wise computations.

## III. RESULTS

### A. MSAM Contribution Analysis

The segmentation performance scores of the baseline networks with different input combinations and with the integration of MSAM for both datasets are shown in Table I. The results for the STS dataset were poorer across the board. In terms of the segmentation performance of U-Net on each type of input combination (CT, PET, or PET-CT) without MSAM, performance using only CT images was especially poor. The scores obtained using only PET were the highest; slightly higher than a PET-CT input.

Tumor segmentation consistently improved with the incorporation of MSAM. For both datasets, the overall top-performing configuration was that which used CT in the backbone with MSAM. This was followed by the two configurations that used PET only: PET in the backbone with MSAM, which surpassed using PET without MSAM.

| | Method | Performance (Mean %) | | | |
|---|---|---|---|---|---|
| | | PREC | SENS | SPEC | DSC |
| Lung Cancer | ResNet-50 (PET-CT) | 71.45 | 74.36 | **99.95** | 67.08 |
| | ResNet-50 (CT) + MSAM | **71.99** | **77.43** | 99.95 | 69.36 |
| | U-Net (CT) | 18.26 | 11.67 | **99.96** | 11.92 |
| | U-Net (PET) | 72.02 | 77.32 | 99.94 | 69.23 |
| | U-Net (PET) + MSAM | 72.51 | 78.54 | 99.94 | 70.01 |
| | U-Net (PET-CT) | 71.78 | 75.19 | 99.95 | 68.22 |
| | U-Net (CT) + MSAM | **72.93** | **81.09** | 99.95 | **71.44** |
| STS | ResNet-50 (PET-CT) | 66.45 | 59.85 | **99.71** | 58.07 |
| | ResNet-50 (CT) + MSAM | **67.54** | **61.89** | 99.71 | **59.59** |
| | U-Net (CT) | 47.90 | 42.23 | 99.70 | 41.35 |
| | U-Net (PET) | 63.45 | **66.09** | 99.55 | 60.19 |
| | U-Net (PET) + MSAM | 66.48 | 64.94 | 99.66 | 61.17 |
| | U-Net (PET-CT) | 64.50 | 64.49 | 99.65 | 59.63 |
| | U-Net (CT) + MSAM | **69.00** | 64.74 | **99.74** | **62.26** |

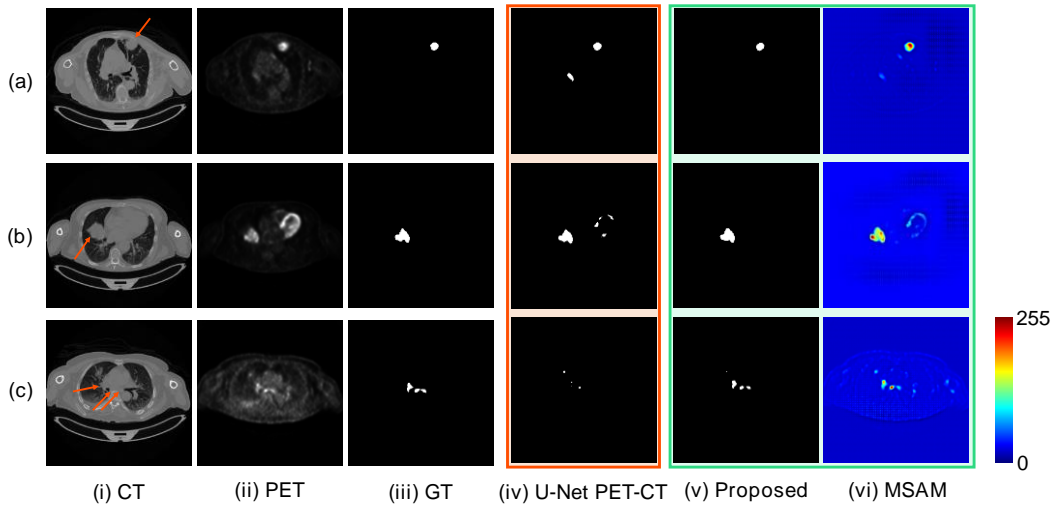

| (i) CT | (ii) PET | (iii) GT | (iv) U-Net PET-CT | (v) Proposed | (vi) MSAM |

Fig. 2. Example outputs from U-Net and our approach for the lung cancer dataset. GT – ground truth segmentation. All images are displayed with a normalized intensity range of 0 to 255.

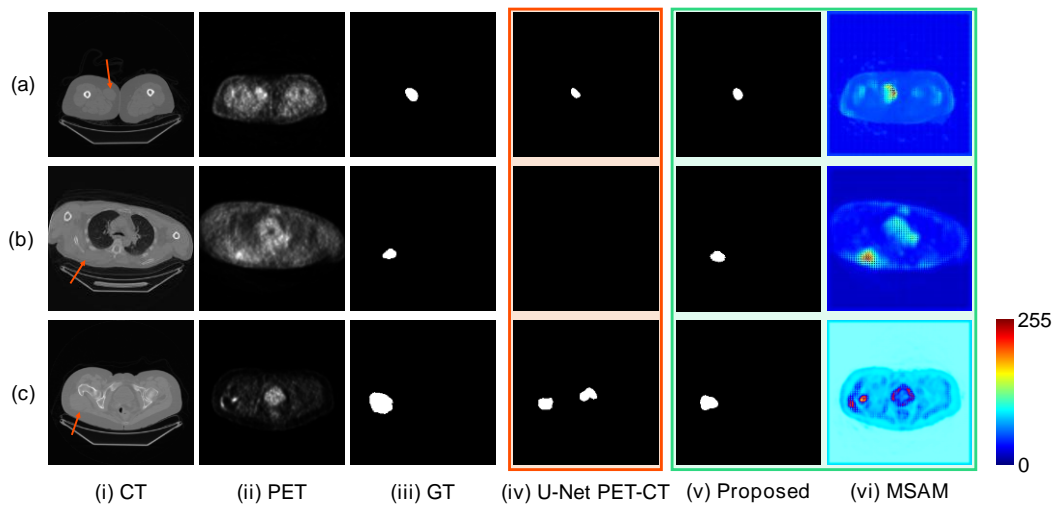

| (i) CT | (ii) PET | (iii) GT | (iv) U-Net PET-CT | (v) Proposed | (vi) MSAM |

Fig. 3. Example outputs from U-Net and our approach for the soft tissue sarcoma dataset. GT – ground truth segmentation. All images are displayed with a normalized intensity range of 0 to 255.

Examining the segmentation outputs of the U-Net PET-CT baseline and the proposed method using MSAM (Fig. 2 and Fig. 3), the baseline U-Net under-segmented (false negative errors) where the uptake of the tumor was relatively small in area or less prominent, or over-segmented (false positive errors) the tumors where the PET image contained non-tumor hotspots e.g., heart in 2(b) and 3(b) or the bladder in 3(c). In contrast, the proposed method avoided such mistakes and produced more accurate segmentations. The MSAM attention maps indicate that bright misleading areas of PET were relatively diminished in intensity compared to tumors.

### B. Comparison Against Current Attention Methods

A comparison of the effects of different attention mechanisms on tumor segmentation performance is presented in Table II. Results for both datasets indicate that MSAM delivered a strong performance boost, while the other methods were inconsistently beneficial, and at times slightly detrimental. MSAM was dominant against the other methods across precision, sensitivity, specificity, and DSC.

TABLE II
COMPARISON OF PERFORMANCE BETWEEN THE MSAM AND STATE-OF-THE-ART ATTENTION APPROACHES.

| Method | | Performance (Mean %) | | | |
|---|---|---|---|---|---|
| | | PREC | SENS | SPEC | DSC |
| Lung Cancer | ARL (PET-CT) | 70.82 | 75.90 | 99.93 | 66.98 |
| | ARL (PET) | 71.09 | 79.95 | 99.93 | 70.07 |
| | CBAM (PET-CT) | 70.80 | 76.16 | 99.94 | 67.83 |
| | CBAM (PET) | 72.66 | 75.67 | 99.94 | 68.71 |
| | AG (PET-CT) | 72.86 | 75.51 | **99.95** | 68.65 |
| | AG (PET) | **74.24** | 78.47 | **99.95** | 71.12 |
| | MSAM | 72.93 | **81.09** | **99.95** | **71.44** |
| STS | ARL (PET-CT) | 64.55 | 62.03 | 99.69 | 58.64 |
| | ARL (PET) | 66.85 | 64.31 | 99.69 | 61.23 |
| | CBAM (PET-CT) | 68.02 | 63.63 | 99.63 | 60.64 |
| | CBAM (PET) | 66.92 | 63.93 | 99.61 | 60.89 |
| | AG (PET-CT) | 64.90 | 62.64 | 99.64 | 58.07 |
| | AG (PET) | 63.59 | 64.35 | 99.64 | 59.67 |
| | MSAM | **69.00** | **64.74** | **99.74** | **62.26** |

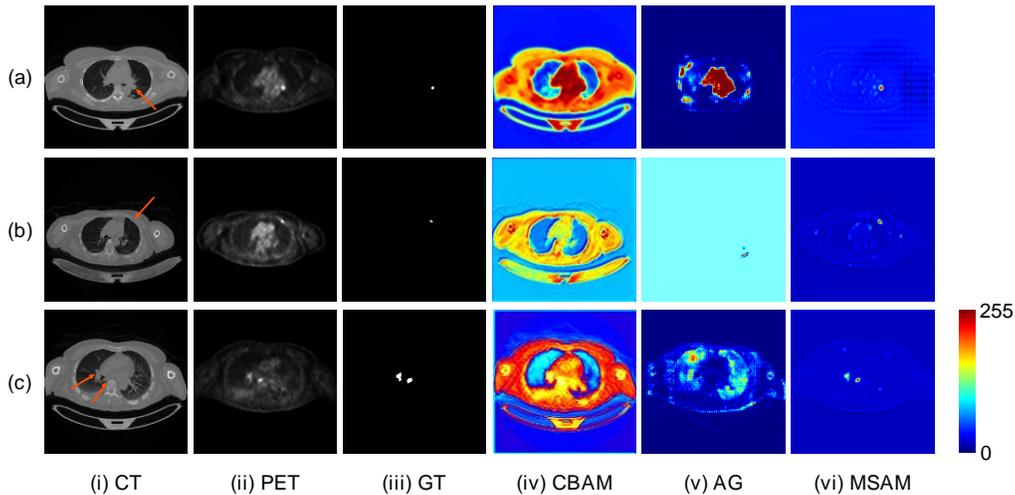

Fig. 4. Comparison of spatial attention maps produced by two current attention methods (CBAM [32] and AG [30]) and MSAM, for transaxial PET-CT images with focal regions of increased FDG uptake in central thoracic lymph nodes. All comparison approaches carried out segmentation on PET-CT. The examples are of the largest resolution maps. All images are displayed with a normalized intensity range of 0 to 255.

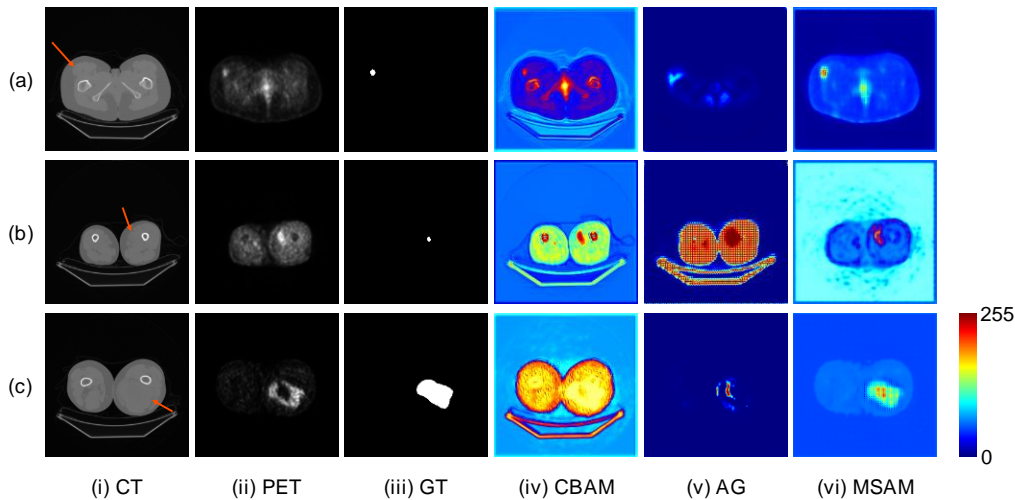

Fig. 5. Comparison of spatial attention maps produced by two current attention methods (CBAM [32] and AG [30]) and MSAM, for transaxial PET-CT images of soft tissue sarcoma tumors. All comparison approaches carried out segmentation on PET-CT. The examples are of the largest resolution maps. All images are displayed with a normalized intensity range of 0 to 255.

The spatial attention maps produced by MSAM indicate that the module clearly intensified tumor regions relative to irrelevant areas, and non-tumor regions were suppressed (Fig. 4 and Fig. 5). In the color heatmaps, pixels belonging to tumors are darker red, while the less important areas are darker blue. In particular, benign areas which display high visual intensity (such as the heart and bladder) in the original PET image were diminished relative to malignant regions, and holes in the tumor areas were filled.

In terms of the other attention methods, the CBAM spatial maps indicate that regions of higher intensity in the CT and PET images were further intensified by this method without regards to whether the region was cancerous. AG sporadically focused on sections of tumors and failed to distinguish non-tumor regions that were brighter in the input PET image. The spatial attention maps produced by MSAM are demonstrably superior at highlighting relevant tumor pixels compared to CBAM and AG.

### C. Comparison Against the State-of-the-Art

| Method | DSC (Mean %) |
| --- | --- |
| Li et al. [24] | 36.45 |
| Zhong et al. [25] | 63.09 |
| Kumar et al. [28] | 63.85 |
| U-Net + MSAM | **71.44** |

A comparison of the proposed MSAM with U-Net against the state-of-the-art is presented in Table III. Our method achieved a mean DSC of 71.44%, which is considerably superior to the other methods; being higher than the previous state-of-the-art by 7.59%.

Segmentation predictions from the various methods for examples of tumors within the lung field and mediastinum are presented in Fig. 6. The methods of Li et al. [24] and Zhong et al. [25] displayed a tendency for under-segmentation, while that of Kumar et al. [28] tended to over-segment the tumors. Our method was able to better capture shape nuances and predict more accurate segmentations.

## IV. DISCUSSION

Our main findings are that: i) MSAM consistently improves the segmentation performance of various backbone CNNs; ii) MSAM is superior at highlighting tumors and enhancing segmentation performance when compared to existing image attention approaches and, iii) our proposed architecture – U-Net backbone + MSAM – outperformed state-of-the-art lung tumor segmentation methods.

### A. MSAM

The original U-Net model without MSAM was our main baseline. Using a channel-concatenated PET-CT input, the model was unable to exploit the strengths of each modality in a complementary manner, as shown by the predicted segmentations (Fig. 2 and Fig. 3), in which there were considerable false positive errors or false negative errors due to an inability to distinguish high-uptake activity as benign or cancerous. These cases indicate that the model incorrectly placed excessive priority on PET and did not proportionately account for corresponding CT information to reduce such errors. The results for single modality input configurations with either PET or CT suggests that PET provides the greatest utility for segmentation, while the poor performance with a CT-only input accentuates the difficulty of tumor segmentation without PET (Table I). Therefore, the concatenation of CT with PET essentially contributed noise to the model and was overall slightly detrimental to performance.

The experiments that involved using only PET images with and without MSAM were effectively PET segmentation (rather than PET-CT). Comparing the performance of the two can help elucidate the effectiveness of the MSAM configuration without effects from CT. The results indicate that MSAM enhances segmentation performance although there was no additional input information provided to the architecture (Table I). This demonstrates the efficacy of the proposed attention approach at helping to focus onto important areas of the input image.

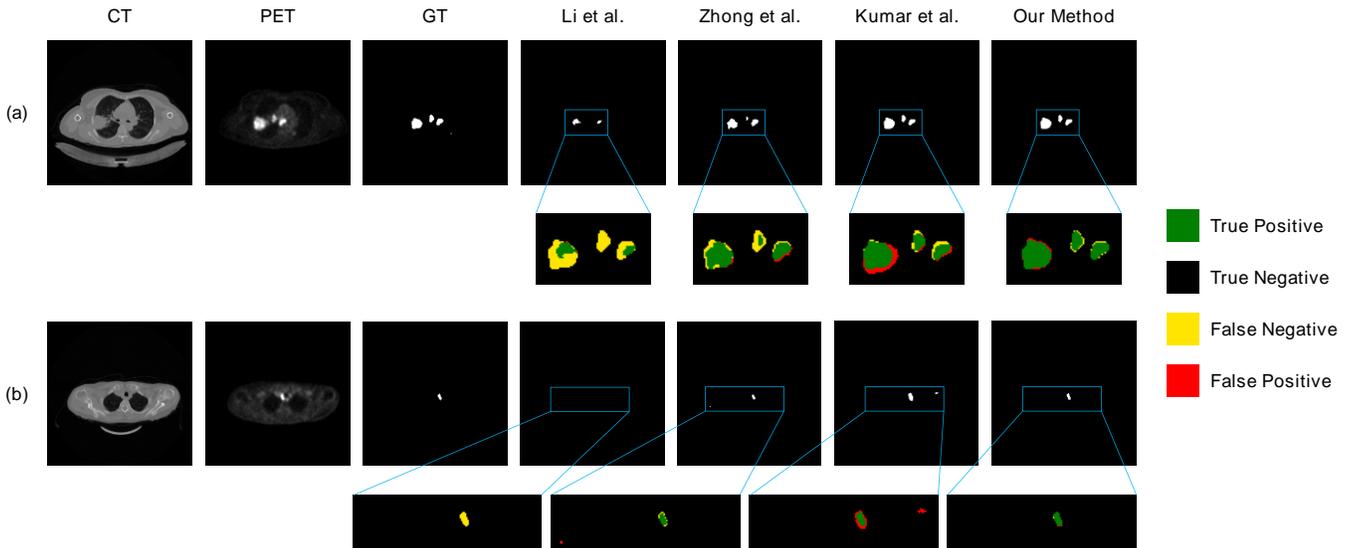

Fig. 6. Visual comparison of the proposed method against previous state-of-the-art at lung tumor PET-CT segmentation. From left to right: CT input image, PET input image, ground truth segmentation, and predictions by Li et al. [24], Zhong et al. [25], Kumar et al. [28], and our method.

We attribute the improvement of our architecture (with CT processed by the backbone and PET by the MSAM) over the PET-only and concatenated PET-CT versions (Table I) due to the configuration in which the MSAM is integrated with the backbone model. This configuration allows a spatial attention map to be learned from PET and used to emphasize corresponding CT features in the more relevant areas. The MSAM was able to filter out misleading hotspots in the PET image such as the heart and bladder (resulting in fewer false positive errors), boost small, less conspicuous, or difficult-to-distinguish areas (avoiding under-segmentation), and fill holes in tumor PET regions (Fig. 2 and Fig. 3). The MSAM learns this automatically in an end-to-end manner without any extra labels or cost functions, as is supported by the way it is integrated with a backbone CNN. Meanwhile in the backbone, the concatenation operations between the attention-weighted and unweighted CT features permit CT information to propagate through the model without effects from the attention map, allowing the model to use information from both modalities in accompaniment. This is exemplified by Fig. 3c, where the attention map only captured fragments of the tumor, but the predicted segmentation was coherent, indicating that CT was used to fill the gaps by supplying morphological guidance. Therefore, our model is not only able to maximize the utility of PET to attend to tumors but also uses the appearance of corresponding locations in CT, thereby exploiting valuable complementary information from both imaging modalities to enhance segmentation.

We have shown that the MSAM can consistently enhance different backbone CNNs for PET-CT segmentation (Table I). Although ResNet-50 is a higher-performing image classifier than the VGG-style encoder of U-Net [41], its tumor segmentation performance was inferior to that of U-Net. A likely reason is from the large loss of resolution due to the initial $7 \times 7$ convolutional layer, which also resulted in the lack of a skip connection at the full lateral resolution of the input. This meant that the full resolution spatial attention map from MSAM was not used for this model. Despite this, segmentation performance was still improved due to the MSAM attention maps applied at lower resolutions.

Overall, tumor segmentation on the STS dataset was evidently more challenging compared to the lung cancer dataset (Table I). The shape and size of the STS tumors were clearly more heterogeneous. Since STS develops in connective tissue, tumors can exist in a diverse range of anatomical locations in the body, such as the legs, trunk, and neck. Consequently, the visual features in the images have greater variation, which increases the difficulty of segmentation. In contrast, in our dataset, the lung tumors were restricted to the lungs.

## B. Comparison Against Current Attention Methods

The MSAM was superior at learning to suppress non-tumor regions (including benign pixels of high intensity) and highlighting tumor regions, compared to existing spatial attention methods. They affected performance inconsistently, as they slightly improved mean DSC on one dataset but was detrimental on the other dataset (Table II).

The spatial attention maps indicate that the CBAM [32] approach only further emphasized the regions that were already brighter in the input, which offers no advantage in terms of spatially constraining the segmentation. The AG [30] approach produced attention maps that were quite random, especially on the lung cancer dataset, where major irrelevant sections were highlighted. The method was at times better at detecting tumor regions on the STS dataset but was unable to distinguish hotspots as tumor or non-tumor (e.g. Fig. 5b). Overall, the results suggest that these benchmark attention methods failed to leverage the high spatial sensitivity for tumors in PET, and the complementary information in the multimodal images. These attention methods lack mechanisms that discern the importance of each modality at different locations as they were not especially designed for multimodal images. In contrast to these existing attention methods, the MSAM helped focus the backbone CNN on regions of higher tumor probability, translating to a consistent boost to segmentation performance.

## C. Comparison Against the State-of-the-Art

In contrast to our MSAM, the previous methods that were compared against employed various fusion strategies to combine information from the two image types. The methods of Li et al. [24] and Zhong et al. [25] tended to under-segment or fail to detect tumors. This was particularly noticeable for tumors in hila (Stage II) or mediastinal (Stage III) lymph nodes (Fig. 6), where the visual features are more variable and difficult to discern due to the adjacent structures, when compared to a tumor in the lung parenchyma where there are fewer surrounding structures. The two methods depend on tumor ROIs to be cropped from the images with the tumor centered. This suggests the requirement of a well-defined area without much variation in functional and anatomical visual features outside the tumor, and a weak ability to identify tumors in the presence of such features. The reduction in performance scores relative to those originally reported can be attributed to these factors. Contrarily, our method does not require tumor boundaries or initialization seeds to be pre-defined, and the results demonstrate that it can handle more varied and challenging anatomical and functional features.

The method of Kumar et al. [28] was more successful at tumor detection but exhibited a tendency for over-segmentation and coarse predictions. This method was primarily concerned about optimal fusion of anatomical and functional visual features between the two image modalities, rather than capturing the more nuanced morphological details that are more critical in segmentation.

## D. Future Work

We have proposed an attention approach for multimodal PET-CT and demonstrated its effectiveness for general backbone CNNs. With the express purpose of illustrating the efficacy of the MSAM configuration, we have not optimized the architecture of the attention module, but rather chose to use a well-established CNN-based model. The architecture may be exchanged for alternative models; refinement of the MSAM architecture is a point of further research. Additionally, the way in which the attention map is applied to the CT features may be improved. Our method uses a conventional elementwise multiplication between the PET attention map and CT features. This is an efficient and intuitive operation, but there may be more advanced procedures that can further enhance segmentation performance.

In our experiments we have used only image slices which contain tumor-positive pixels. Identification of such slices require a tumor detection pre-processing step that relies on manual input. To make the process more streamlined and applicable to a wider range of datasets, an initial automatic tumor detection framework may be applied to extract slices that contain tumors.

The images in our experiments were processed as 2D slices rather than as 3D volumes. 3D CNNs may deliver superior performance, albeit with a much greater computational expense, penalty to speed, and increased number of trainable parameters with a higher risk of overfitting. The extension of the MSAM to a 3D framework is a possible avenue for further investigation.

We have used two independent datasets with a combined total of 100 PET-CT scans on two different cancer types. This is a substantial amount of experimental data relative to comparable studies [24, 25, 28]. We will further refine our approach and evaluate among other cancer datasets.

## V. Conclusion

We proposed a spatial attention approach to improve the performance of CNNs at segmenting tumors from multimodal PET-CT images. Our MSAM automatically learns to spatially emphasize tumor regions while suppressing benign areas. When integrated with backbone CNNs, the MSAM substantially enhances segmentation performance, and is superior to existing image spatial attention methods. Our approach surpassed previous state-of-the-art methods for PET-CT lung tumor segmentation.